\documentclass[review]{elsarticle}

\usepackage{xcolor}
\usepackage{subcaption}
\usepackage{lineno,hyperref}
\usepackage{latexsym}
\usepackage{tabularx}
\usepackage{multirow}
\usepackage{multicol}
\usepackage{caption}
\usepackage{booktabs}
\modulolinenumbers[5]
\usepackage{float}
\floatstyle{plaintop}
\restylefloat{table}

\journal{Data in Brief}

\begin{document}

\begin{frontmatter}

\title{ContextLabeler Dataset: physical and virtual sensors data collected from smartphone usage in-the-wild\tnoteref{mytitlenote}}
\tnotetext[mytitlenote]{this work is co-submitted with the paper entitled: "COMPASS: an Unsupervised and Online solution for Context-aware mobile applications".}

\author[iitaddress]{Mattia G. Campana\footnote{Mattia G. Campana is corresponding author, National Research Council of Italy IIT-CNR, mattia.campana@iit.cnr.it}}
\author[iitaddress]{Franca Delmastro}

\address[iitaddress]{Institute of Informatics and Telematics, National Research Council of Italy, Pisa, Italy \\mattia.campana@iit.cnr.it, franca.delmastro@iit.cnr.it}

\begin{abstract}
This paper describes a data collection campaign and the resulting dataset derived from smartphone sensors characterizing the daily life activities of 3 volunteers in a period of two weeks.
The dataset is released as a collection of CSV files containing more than 45K data samples, where each sample is composed by 1332 features related to a heterogeneous set of physical and virtual sensors, including motion sensors, running applications, devices in proximity, and weather conditions.
Moreover, each data sample is associated with a ground truth label that describes the user activity and the situation in which she was involved during the sensing experiment (e.g., \emph{working}, \emph{at restaurant}, and \emph{doing sport activity}).
To avoid introducing any bias during the data collection, we performed the sensing experiment in-the-wild, that is, by using the volunteers' devices, and without defining any constraint related to the user's behavior.
For this reason, the collected dataset represents a useful source of real data to both define and evaluate a broad set of novel context-aware solutions (both algorithms and protocols) that aim to adapt their behavior according to the changes in the user's situation in a mobile environment.
\end{abstract}


\end{frontmatter}


\section{Specifications Table}
\begin{table}[H]
    \scriptsize
    \centering
    \begin{tabularx}{\textwidth}{lX}
    \toprule
        \textbf{Subject} & Computer Science \\
        \textbf{Specific subject area} & User context sensing and modeling for detecting human complex activities and situations in mobile environments.\\
        \textbf{Type of data} &  Comma Separated Files (CSV).\\
        \textbf{How data were acquired} & Android mobile application.\\
        \textbf{Data format}	& Raw sensors data with user annotations.\\
        \textbf{Parameters for data collection} & The experiment has been designed to collect context data into the wild. In other words, to avoid introducing biases during the data acquisition, we did not define any constraints for the user behavior during the experiment. For example, we encouraged the volunteers to use their smartphones without worrying about the positions of the device (e.g., trousers’ pockets, or hand).\\
        
        \textbf{Description of data collection} & In order to take into account the diversity of different devices, the volunteers have installed the sensing application on their smartphones. The mobile application has been designed for Android OS, and it collects data generated by a heterogeneous set of sensors, including both physical and virtual sensors. The collected data was stored in the internal storage unit of the mobile device.
        
        The volunteers were able to start and stop the sensing application whenever they wanted, and they freely annotated the collected data by choosing among a set of predefined daily life activities.\\
        \textbf{Data source location} & Institution: National Research Council of Italy, Pisa, Italy
        City/Town/Region: Pisa, Lucca and Pontedera (i.e., 3 different towns in Tuscany region, Italy)
        Country: Italy
        \\
        \textbf{Data accessibility} & Direct URL to data: \url{https://github.com/contextkit/ContextLabeler-Dataset}\\
        \textbf{Related research article} & M. G. Campana, D. Chatzopoulos, F. Delmastro, and P. Hui. 2018. Lightweight Modeling of User Context Combining Physical and Virtual Sensor Data. In Proceedings of the 2018 ACM International Joint Conference and 2018 International Symposium on Pervasive and Ubiquitous Computing and Wearable Computers (UbiComp ’18). Association for Computing Machinery, New York, NY, USA, 1309–1320.
        DOI: https://doi.org/10.1145/3267305.3274178\\
        \bottomrule
    \end{tabularx}
    \label{tab:specification}
\end{table}{}

\section{Value of the Data}
%
\begin{itemize}

    \item The presented dataset provides a broad set of sensors data describing human complex activities collected from the use of commercial smartphones into-the-wild. All the data samples have been freely annotated by the users in order to specify their daily life activities. Moreover, since we have not defined any constraints for the user behaviour, the presented dataset is not affected by biases that can be introduced by performing predefined actions in controlled environments (e.g., laboratory).
    
    \item Researchers can use this dataset to analyze and automatically recognize the situation in which the user is currently involved by using commercial smartphones.
    
    \item This dataset provides a valuable starting point for the automatic detection of the user context in a mobile setting. Specifically, it can be used to evaluate novel context-aware solutions, including recommender systems, activity recognition algorithms, and wireless communication protocols.
    
    \item The dataset presents also additional values: (i) the data has been collected in real environments, without defining any sort of constraints related to the user behaviour nor to the interactions between the user and her mobile device; (ii) each data sample is represented by a high-dimensional vector composed by more than 1K features extracted from a heterogeneous set of mobile sensors (both physical and virtual); (iii) the data has been freely annotated by the users according to their daily life activities.

\end{itemize}
\section{Data Description}

The dataset contains smartphone sensors data collected from the personal devices of 3 volunteer users in their usual environment.
It is released in the form of a set of comma-separated (CSV) files, one for each volunteer, and they are respectively named as follows: \texttt{user\_1.csv}, \texttt{user\_2.csv}, and \texttt{user\_3.csv}.
The CSV files contain time series of sensors data collected from the users' devices through a mobile application specifically designed for the sensing experiment.
This application has been used by the volunteers for two weeks to annotate the collected sensed data with labels that describe their daily life activities. In total, we collected 45681 data samples that are contained in the three files as follows: 8456 samples in \texttt{user\_1.csv}, 17882 samples in \texttt{user\_2.csv}, and 19343 in \texttt{user\_3.csv}.

The dataset contains both physical and virtual sensors data that can be used to characterize all the different aspects of the user context in a mobile setting.
Physical sensors are implemented in the hardware equipment of the mobile phone (e.g., accelerometer), while virtual sensors represent data sources that describe the device's status, the surrounding environment, and the interactions between the user and her device.

Each data sample is composed by 1332 features, with both continuous and categorical values, describing a heterogeneous set of sensors.
According to the type of sensors they describe, we can divide the collected features in the following 13 categories:

\begin{description}
    \item[Date and Time]: each data sample is associated with a Unix timestamp that represents the instant in which our sensing application has captured the sensors data. Starting from the timestamp, we also extracted 5 categorical features related to both day and time, i.e., \emph{weekday}, \emph{weekend}, \emph{morning}, \emph{afternoon}, \emph{evening}, and \emph{night}.
    
    \item[User gait]: 8 categorical features that represent the user's gait detected by the Android Activity Recognition API~\footnote{\url{https://developers.google.com/android/reference/com/google/android/gms/location/DetectedActivity.html}}:
    \begin{itemize}
        \item \emph{activity\_rec\_in\_vehicle}: the user is in a vehicle (e.g., a car),
        \item \emph{activity\_rec\_on\_bicycle}: the user is riding a bicycle,
        \item \emph{activity\_rec\_on\_foot}: the user is walking or running,
        \item \emph{activity\_rec\_running}: the user is running,
        \item \emph{activity\_rec\_still},
        \item \emph{activity\_rec\_tilting}, the user is rapidly moving the device,
        \item \emph{activity\_rec\_walking}, the user is walking,
        \item \emph{activity\_rec\_unknown}, the Google API is not able to recognize the current user's activity
    \end{itemize}
    
    \item[Running applications]: 56 categorical features that represent the possible main application categories, according to the Google Play Store (e.g., \emph{ART\_AND\_DESIGN}, \emph{BUSINESS}, and \emph{ENTERTAINMENT}).
    The value of a feature represents the number of running applications that belong to the corresponding application category.
    
     \item[Weather conditions]: based on the user's location, we defined a total of 62 features by using the information collected from the OpenWeatber API service~\footnote{\url{https://openweathermap.org/api}}.
     More specifically, we defined the following 8 continuous features:
     \begin{itemize}
        \item \emph{weather\_temp}: the current temperature in Celsius,
        \item \emph{weather\_temp\_min}: the minimum temperature of the day,
        \item \emph{weather\_temp\_max}: the maximum temperature of the day,
        \item \emph{weather\_humidity}: the percentage of humidity,
        \item \emph{weather\_pressure}: the atmospheric pressure in hPa,
        \item \emph{weather\_wind\_speed}: the wind speed in meter/sec,
        \item \emph{weather\_wind\_direction}: the wind direction in degrees,
        \item \emph{weather\_cloudiness}: the percentage of cloudiness
     \end{itemize}
     In addition, we defined a total of 54 categorical features derived from the weather conditions codes defined by the OpenWeather service~\footnote{\url{https://openweathermap.org/weather-conditions}}.
     
     \item[Audio]: a set of 12 categorical and continuous features related to the current smartphone’s audio settings. Specifically, we defined 4 categorical features to represent the ringer mode (i.e.,  \emph{audio\_ringer\_mode\_silent}, \emph{audio\_ringer\_mode\_vibrate}, and \emph{audio\_ringer\_mode\_normal}), and the following 5 categorical features for other audio characteristics: \emph{audio\_bt\_sco\_on} and \emph{audio\_headset\_on}, that respectively represent whether a bluetooth and a wired headset is connected to the device or not; \emph{audio\_music\_active} and \emph{audio\_speaker\_on} that respectively indicate if the music and the speaker are active; and \emph{audio\_mic\_mute}, that represent if the microphone is on or off.
     In addition, we defined the following continuous features to characterize the level of different audio settings:
     \begin{itemize}
        \item \emph{audio\_alarm\_volume}: the alarm volume,
        \item \emph{audio\_music\_volume}: music volume,
        \item \emph{audio\_notification\_volume}: the volume level set for the notifications,
        \item \emph{audio\_ring\_volume}: the ringtone volume
     \end{itemize}
     
     \item[Battery]: 4 categorical features related to the battery information. Specifically, a feature that represents whether the device is connected to a power source or not (i.e., \emph{battery\_unplugged}), and 3 features to characterize the type of power source: \emph{battery\_plugged\_ac} (an AC charger), \emph{battery\_plugged\_usb} (a USB port), and \emph{battery\_plugged\_wireless} (an inductive wireless charger).
     
     \item[Bluetooth connections]: 54 categorical features that characterize the type of the first 5 Bluetooth devices connected to the user's smartphone. Specifically, based on the Bluetooth Major ID number, the possible device categories are the following: \emph{audio\_video}, \emph{computer}, \emph{health}, \emph{imaging}, \emph{misc}, \emph{networking}, \emph{peripheral}, \emph{phone}, \emph{toy}, \emph{wearable}, and \emph{uncategorized}.
     
     \item[Bluetooth devices in proximity]: 54 categorical features that characterize the type of the first 5 Bluetooth devices in proximity.
     
     \item[Display status]: 11 categorical features that represent information about the display status. Specifically, the following features describe the current display state:
     
     \begin{itemize}
        \item \emph{display\_status\_on}: the display is on,
        \item \emph{display\_status\_off}: the display is off,
        \item \emph{display\_status\_doze}: the display is in a low-power state: the display shows only system-provided content while the device is non-interactive,
        \item \emph{display\_status\_doze\_suspended}: the display is in a suspended low-power state, where the CPU is no more updating it,
        \item \emph{display\_status\_vr\_mode}: the display is optimized for the Virtual Reality (VR) mode,
        \item \emph{display\_status\_on\_suspended}: the display is in a full-power mode, but the display is not updating it,
        \item \emph{display\_status\_unknown}: the Android system is not able to recognize the current display status,
     \end{itemize}
     
     while the following features characterize the rotation angle of the display: \emph{display\_rotation\_0} (natural -vertical- orientation), \emph{display\_rotation\_90} (horizontal mode), \emph{display\_rotation\_180} (vertical and rotated by 180 degree), and \emph{display\_rotation\_270} (horizontal and rotated by 270 degree).
     
     \item[Location]: two continuous features that respectively represent the geographical coordinates (i.e., latitude and longitude) of the user's current location. Moreover, based on the user's location, we downloaded the category of the most probable venue according to the Foursquare Places API (e.g., Art Gallery or Italian Restaurant)~\footnote{\url{https://developer.foursquare.com/docs/api/venues/search}}. Therefore, we also defined 921 categorical features that represent the main venue categories defined by Foursquare~\footnote{\url{https://developer.foursquare.com/docs/resources/categories}}.
     
     \item[Wi-Fi]: a categorical feature that represents whether the mobile device is currently connected to a Wi-Fi Access Point or not.
     
     \item[Physical Sensors]: a set of 136 continuous features that represent several descriptive statistics related to the following physical sensors: light, accelerometer, gravity, gyroscope, linear acceleration, rotation, and proximity.
     More specifically, for each sensor we collected 200 data samples and we calculated the following statistics: minimum, maximum, and average values; quadratic mean; 25th, 50th, 75th, and 100th percentiles.
     Moreover, for those sensors that are composed of multiple components (e.g., a 3-axis gyroscope), we calculated the same set of statistics for each component.
     
     \item[Multimedia]: 2 categorical features that represent whether the user was taking a picture or recording a video with her smartphone.

\end{description}

Finally, each data sample is associated with its Ground Truth: the label specified by the user to describe the type of situation in which she was involved when the application collected the sensors' data.
The labels specified by the users are the following: \emph{Working}, \emph{Restaurant}, \emph{Launch Break}, \emph{Shopping}, \emph{Break}, \emph{Home}, \emph{Nightlife}, \emph{Sleep}, \emph{Physical exercise}, and \emph{Free time}.

\section{Experimental Design, Materials, and Methods}

The dataset we release with this work is the result of a data collection campaign designed to capture the complexity of the user context in a mobile environment.
With the term context, here we mainly refer to the activities performed by a person during her daily life and the situations in which she can be involved.
Examples of possible contexts are the following: \emph{attending to a lecture}, \emph{being at home}, and \emph{taking a coffee with friends}.

According to the literature~\cite{8430551}, simple human activities (e.g., \emph{running} or \emph{walking}) can be characterized by using a small set of sensors embedded in personal and wearable devices like, for example, the accelerometer and the gyroscope.
On the contrary, complex activities have higher-level semantics and require a combination of heterogeneous sources of data.
Therefore, to infer the user situation by using the sensing capabilities of her mobile and personal devices, we need to take into account a broad set of heterogeneous sources of data.
To this aim, the simple physical sensors are not enough, but we also need to exploit the so-called virtual sensors, i.e., those data sources that characterize the user-device interactions as long as the surrounding environment (e.g., running applications and devices in proximity).

Research studies in the area of activity recognition and human behavior modeling usually base their results on experiments performed in controlled environments (e.g., a research laboratory)~\cite{micucci2017unimib}.
During the data collection process (often performed with the same device), volunteers are asked to perform some activities that have been previously defined by researchers.
However, in the real world, we have heterogeneous devices and different users may have different ways of doing the same activity; thus the experimental results usually diverge from those obtained in the lab~\cite{mafrur2015modeling}.

To build a realistic and valuable dataset, we enrolled three voluntary users equipped with heterogeneous commercial mobile devices, with different characteristics and sensors: a Nexus 5 with Android 6.0.1, a Xiaomi Mi 5 with Android 7.1.2, and a Reader P10 with Android 6.0. 
Users signed an informed consent including all the policies adopted for personal data storage, management, and analysis, including the publication of the anonymized dataset, according to the EU GDPR.
Besides, to avoid introducing biases during the data acquisition, we did not define any constraints for the user behavior during the experiment. On the contrary, we encouraged the volunteers to use their smartphones without worrying about the positions of the device (e.g., trousers’ pockets, or hand).

\begin{figure}[t]
    \begin{subfigure}{.49\textwidth}
        \centering
        \includegraphics[width=.6\linewidth]{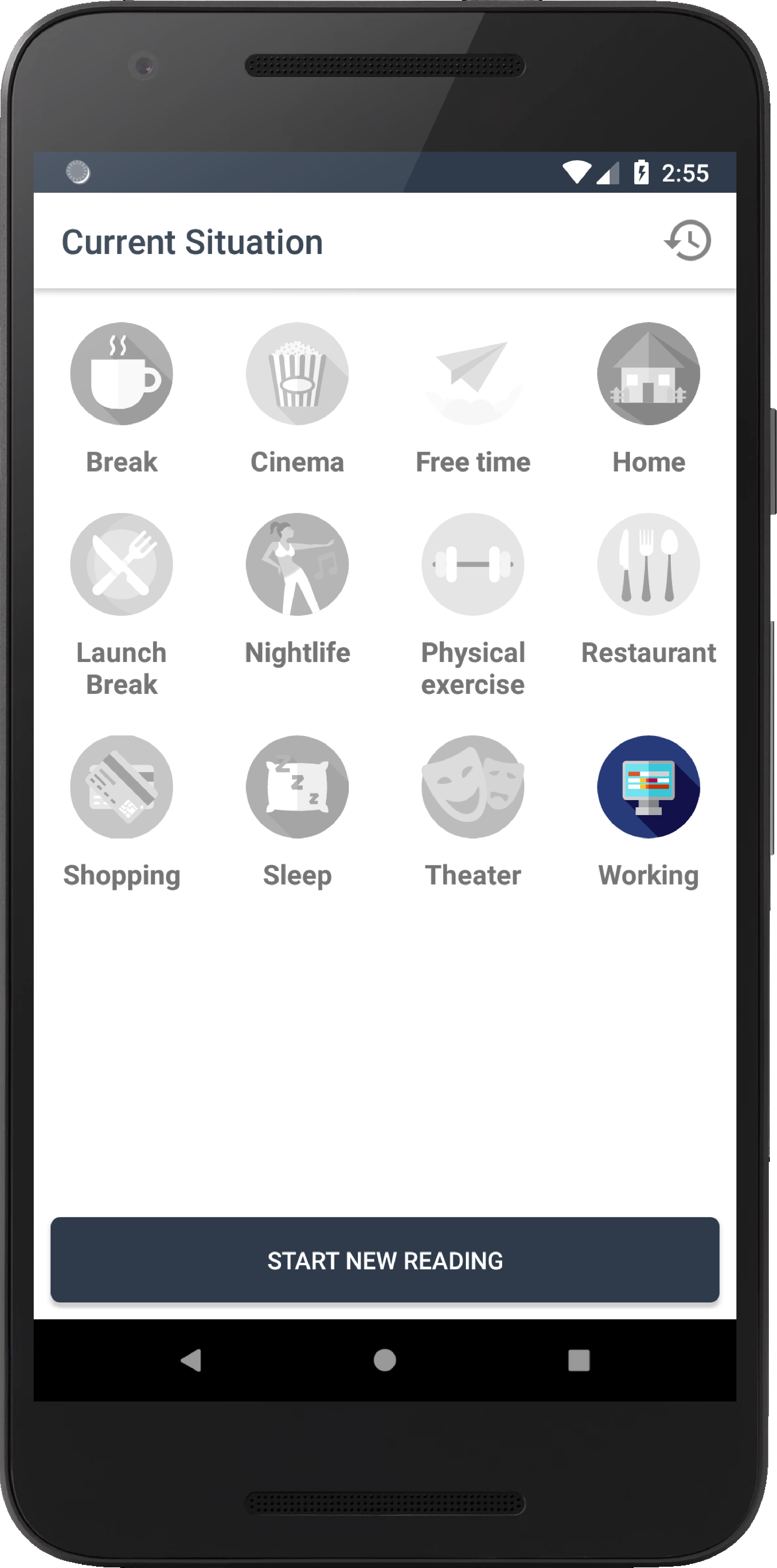}
        \caption{Label selection.}
        \label{fig:cl_labels}
    \end{subfigure}
    \begin{subfigure}{.49\textwidth}
        \centering
        \includegraphics[width=.6\linewidth]{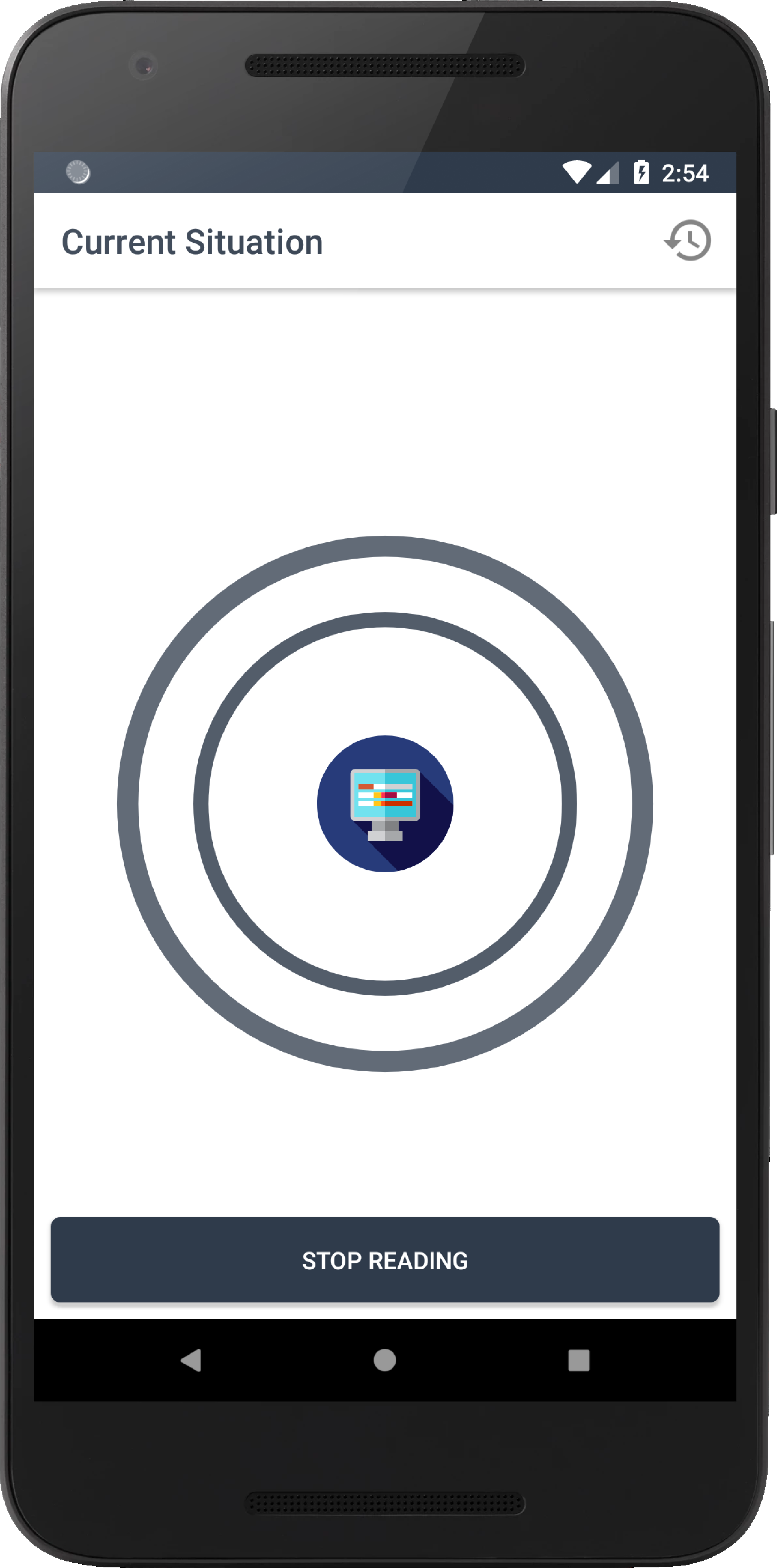}
        \caption{Reading sensors.}
        \label{fig:cl_reading}
    \end{subfigure}
    \caption{User Interface of the Context Labeler mobile application.}
\end{figure}

To collect the dataset we developed Context Labeler, an Android application that allows the volunteers to freely annotate the sensed data.
More specifically, we asked the volunteers to install the sensing application on their smartphones and to select their daily life activities among the following set of labels: \emph{Break}, \emph{Cinema}, \emph{Free time}, \emph{Home}, \emph{Lunch Break}, \emph{Nightlife}, \emph{Physical exercise}, \emph{Restaurant}, \emph{Shopping}, \emph{Sleep}, \emph{Theatre}, and \emph{Working}.
Figure~\ref{fig:cl_labels} shows the User Interface offered by Context Labeler to specify the current user's context.
After the activity selection, Context Labeler starts ContextKit~\footnote{\url{https://contextkit.github.io/}}, our sensing framework that monitors a broad set of sensors, both physical and virtual sensors~\cite{10.1145/3267305.3274178}.
In order to avoid affecting the user behavior and the interactions with her device, the data collection is completely performed unobtrusively in the background.
When the current activity ends, the user manually stops the data reading using a specific button (Figure~\ref{fig:cl_reading}), and both the sensed data and the selected label are stored into the device’s hard drive.

\begin{table}[ht!]
\centering
\caption{Sensors categories and their sampling rates used during the data collection.}
\begin{tabular}{lr}
    \toprule
    \textbf{Category} & \textbf{Sampling rate (sec.)}
    \\ \midrule
    User gait                   & 60\\
    Running applications        & 300\\
    Weather conditions          & 3600\\
    Audio                       & 60\\
    Battery                     & 60\\
    Bluetooth connections       & on event\\
    Bluetooth scans             & 60\\
    Display                     & 60\\
    Location                    & 300\\
    Wi-Fi                       & 180\\
    Physical sensors            & 60\\
    Multimedia                  & on event
    \\\bottomrule
\end{tabular}
\label{tab:sampling_rate}
\end{table}

Table~\ref{tab:sampling_rate} shows the sampling rate of each sensor's category that we have used in Context Labeler during the data collection.
When the user starts the sensing procedure, the application collects new data samples every 1-5 minutes for most of the sensors; while it downloads the weather conditions every hour from the OpenWeather service.
Moreover, both the \emph{Bluetooth Connections} and the \emph{Multimedia} sensors react to specific events.
Specifically, when the user connects or disconnects a Bluetooth device to her smartphone, or she takes a photo or records a video, Context Labeler saves such information on the log files.

ContextKit stores the sensed data in dedicated log files, one for each monitored sensor, alongside with the reading timestamps.
However, different sensors or events monitored by the framework may have different sampling rates. Therefore, even if two different sensor data refer to the same user context, they may have slightly different timestamps. Moreover, each label collected by the application is stored, together with its duration, in a separate log file.

\begin{figure}[t]
    \centering
    \includegraphics[width=.6\textwidth]{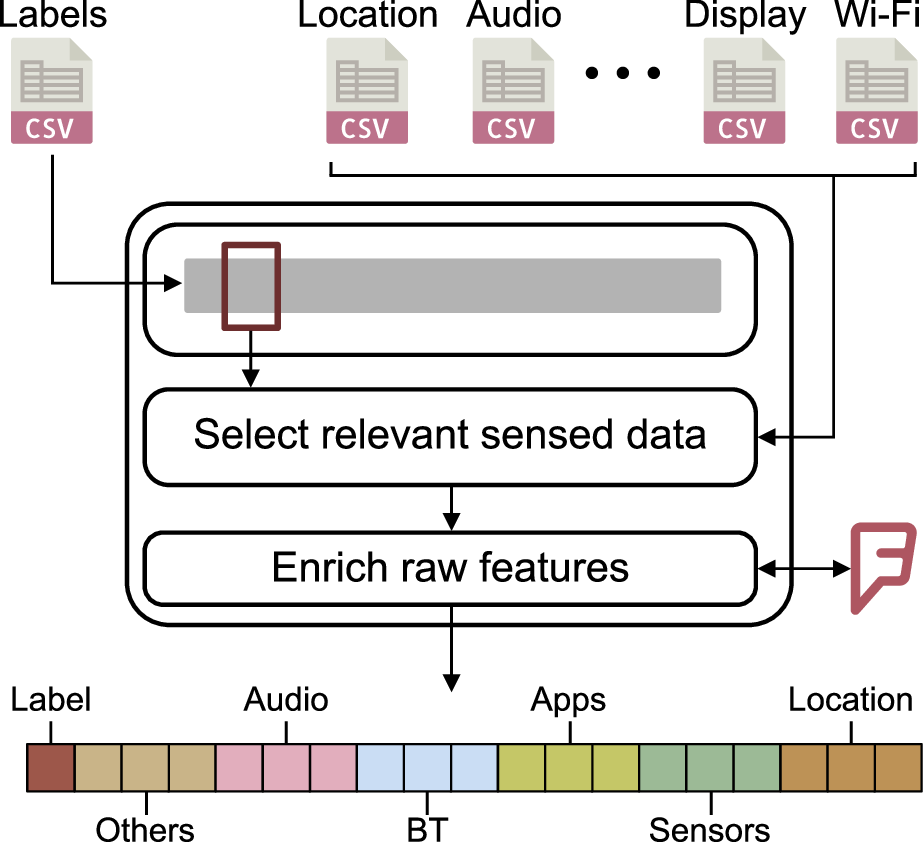}
    \caption{Dataset building process.}
    \label{fig:dataset_building_process}
\end{figure}

In order to generate a dataset which is ready to be used for research purposes (e.g., to evaluate context-recognition algorithms), we processed the log files as shown in Figure~\ref{fig:dataset_building_process}.
First, we used a sliding window approach to split the duration of each user situation into slots of 1 second each.
Second, for every time slot, we fetched from the raw log files only the sensor data with the closest reading timestamp to the starting time of the current slot.
In this way, we kept only dense feature vectors, and we discard data samples wit missing values.
Then, we enriched the raw features with additional categorical information.
For example, using the \emph{Foursquare APIs}~\footnote{\url{https://developer.foursquare.com}}, we extended the location features by retrieving the category of the most probable venue according to the GPS coordinates.
Finally, we have created the final feature vector by combining the categorical features with the continuous ones derived from physical sensors values, and we associate the corresponding situation's label indicated by the user.

Since an ordinal relationship among categorical features does not exist, we include the categorical features into the features vector by using the well-known One Hot Encoding technique, which creates a binary feature for each possible category.
For example, according to the Android Framework~\footnote{\url{https://developer.android.com/reference/android/view/Display\#getOrientation()}}, the possible display orientation modes are the following: 0, 90, 180, and 270 degrees.
Assuming that, for a given context snapshot, the display was held by the user in portrait mode, we have created 4 different features for describing the display status, where one of them (i.e., the one corresponding to 0 degrees) is set to 1, while the others are set to 0.
The resulting dataset contains 45681 labeled samples, where each sample is composed of 1332 features.

\section*{Acknowledgments}
This work has been carried out in the framework of the INTESA project (CUP CIPE D78I16000010008), co-funded by the Tuscany Region (Italy) and MIUR under the Programme FAR FAS 2007-2013.

\section*{Competing Interests}
The authors declare that they have no known competing financial interests or personal relationships that could have appeared to influence the work reported in this paper.


\bibliographystyle{elsarticle-num}
\bibliography{main}

\end{document}